\documentclass[pdflatex,sn-nature]{sn-jnl}

\usepackage{graphicx}%
\usepackage{multirow}%
\usepackage{amsmath,amssymb,amsfonts}%
\usepackage{amsthm}%
\usepackage{mathrsfs}%
\usepackage[title]{appendix}%
\usepackage{xcolor}%
\usepackage{textcomp}%
\usepackage{manyfoot}%
\usepackage{booktabs}%
\usepackage{algorithm}%
\usepackage{algorithmicx}%
\usepackage{algpseudocode}%
\usepackage{listings}%

\theoremstyle{thmstyleone}%
\theoremstyle{thmstyletwo}%

\theoremstyle{thmstylethree}%

\begin{document}

\title[Article Title]{\bfseries Extreme Terahertz Nonlinear Phononics by Coherence-Imprinted Control of Hybrid Order}

\author*[1]{\fnm{Liang} \sur{Luo}}\email{liangluo@ameslab.gov (L.L.)}


\author[1,2]{\fnm{Avinash} \sur{Khatri}}

\author[1]{\fnm{Martin} \sur{Mootz}}

\author[1]{\fnm{Tao} \sur{Jiang}}

\author[3,4]{\fnm{Liu} \sur{Yang}}

\author[5,6]{\fnm{Zijing} \sur{Chen}}

\author[1,2]{\fnm{Chuankun} \sur{Huang}}

\author[1,2]{\fnm{Zhi Xiang} \sur{Chong}}

\author[1]{\fnm{Joongmok} \sur{Park}}

\author[7]{\fnm{Ilias E.} \sur{Perakis}}

\author[3,4,8]{\fnm{Zhiwei} \sur{Wang}}

\author[3,4,8]{\fnm{Yugui} \sur{Yao}}

\author[5,6]{\fnm{Dao} \sur{Xiang}}

\author[1,2]{\fnm{Yong-Xin} \sur{Yao}}

\author*[1,2]{\fnm{Jigang} \sur{Wang}}\email{jgwang@ameslab.gov (J.W.)}

\affil[1]{\fontsize{10}{20}\selectfont\orgdiv{Ames National Laboratory}, \orgname{U.S. Department of Energy}, \city{Ames}, \state{IA}, \postcode{50011}, \country{USA}\vspace*{.4em}}

\affil[2]{\fontsize{10}{17}\selectfont\orgdiv{Department of Physics and Astronomy}, \orgname{Iowa State University}, \city{Ames}, \state{IA}, \postcode{50011}, \country{USA}\vspace*{.4em}}

\affil[3]{\fontsize{10}{17}\selectfont\orgdiv{Centre for Quantum Physics, Key Laboratory of Advanced Optoelectronic Quantum Architecture and Measurement (MOE), School of Physics}, \orgname{Beijing Institute of Technology}, \city{Beijing}, \postcode{100081},  \country{China}\vspace*{.4em}}

\affil[4]{\fontsize{10}{17}\selectfont\orgdiv{Beijing Key Lab of Nanophotonics and Ultrafine Optoelectronic Systems}, \orgname{Beijing Institute of Technology}, \city{Beijing}, \postcode{100081},  \country{China}\vspace*{.4em}}

\affil[5]{\fontsize{10}{17}\selectfont\orgdiv{Key Laboratory for Laser Plasmas (Ministry of Education), School of Physics and Astronomy}, \orgname{Shanghai Jiao Tong University}, \city{Shanghai}, \postcode{200240},  \country{China}\vspace*{.4em}}

\affil[6]{\fontsize{10}{17}\selectfont\orgdiv{Zhangjiang Institute for Advanced Study and Tsung-Dao Lee Institute}, \orgname{Shanghai Jiao Tong University}, \city{Shanghai}, \postcode{201210},  \country{China}\vspace*{.4em}}

\affil[7]{\fontsize{10}{17}\selectfont\orgdiv{Department of Physics}, \orgname{University of Alabama at Birmingham}, \city{Birmingham}, \state{AL}, \postcode{35294}, \country{USA}\vspace*{.4em}}

\affil[8]{\fontsize{10}{17}\selectfont\orgname{Beijing Institute of Technology}, \city{Zhuhai}, \postcode{519000},  \country{China}\vspace*{.4em}}


\abstract{\normalsize
Coherent control of quantum materials has progressed along two major fronts: nonlinear phononics, which reshapes lattices to induce emergent states, and Floquet engineering, which tailors electronic band reconstruction via time-periodic driving. Both mechanisms face fundamental limitations at terahertz (THz) frequencies: phononic nonlinearities are intrinsically weak in standard lattices, while electronic Floquet states are often constrained by rapid decoherence upon light-off and by a scarcity of coherence-resolved, multi-correlation probes beyond (quasi-)stationary band structures. Here we report an extreme THz nonlinear-phononics mechanism in $\text{Ta}_\text{2}\text{NiSe}_\text{5}$, where a highly susceptible non-equilibrium electronic correlation bath dramatically amplifies lattice nonlinearities under coherent driving. Utilizing THz two-dimensional spectroscopy as a coherence-tomography tool, we resolve an exceptionally rich landscape of approximately 30 distinct multi-order quantum pathways, including high-harmonic phonon generation, multi-quantum coherences, and multi-wave anharmonic cross-mode mixing.
The density and complexity of this extreme manifold establishes a new benchmark for THz nonlinear phononics, as the multi-order quantum pathways surpass the limits of conventional lattice responses. These high-order signals collapse above $\sim$100~K, defining an electronic correlation scale of a coherence-imprinted hybrid electronic-phonon order that governs the sustainability of high-order quantum correlations and nonlinear pathways beyond linear and equilibrium responses. Our results establish a route for correlation-boosted, phonon-anchored periodic Hamiltonian engineering and for certifying such periodically-driven states via multi-correlation coherence tomography.}

\maketitle

\subsubsection*{Introduction}\label{sec1}

The lattice framework underpins a wide range of emergent phenomena by shaping electronic, magnetic, optical, and chiral degrees of freedom--from high-temperature superconductivity~\cite{MankowskyNature2014} and magnetism~\cite{Rini2007,LiNature2013, Afanasiev2021} to ferroelectricity~\cite{Li2019,Mankowsky2017,Kozina2019} and topological order~\cite{Sie2019,LuoNatureMat2021,VaswaniPRX2020, YangNPJQM2020}.
{\em Nonlinear phononics} now allows selective driving of coherent lattice vibrations with intense terahertz (THz) fields, coherently manipulating the properties of quantum materials on ultrafast timescales~\cite{MankowskyNature2014, subedi2014theory,Ghalgaoui2025,Blank2023,Kozina2019,VaswaniPRX2020}. Yet, compared with the well-studied and strong nonlinearities of electrons and magnons~\cite{HuangNC2024,Lu2017,Hafez2018,LiNature2013,Rohrbach2025}, phononic nonlinear responses are intrinsically weaker and have been far less explored~\cite{Forst2011,Disa2021,vonHoegen2018}. 
Conventional single-axis spectroscopy, as used in most prior demonstrations~\cite{Disa2021,vonHoegen2018}, resolves only harmonic generation and thus misses the multi-quantum phonon coherences and anharmonic mixing among distinct phonon modes along the excitation-emission plane~\cite{HuangNRP2025}. The latter, extreme form of nonlinear phononics arises from exceptionally rich nonlinear quantum pathways and multi-order correlation functions that remain hidden thus far. Access to this multidimensional landscape via coherence-resolved multi-correlation tomography~\cite{HuangNC2024,HuangNRP2025,ZhangNatPhys2024, Rohrbach2025} is crucial for uncovering electronically enhanced phononic responses and for integrating them with coherence-imprinted, periodic Hamiltonian engineering principles~\cite{McIverNP2020,RechtsmanNat2013,SZhouNat2023,YJShanNat2021,SItoNat2023,Henstridge2022,MBorschSci2020} to achieve unified coherent control.
This coherence-imprinted, hybrid electron--phonon response regime overcomes both limits: weak phononic nonlinearities and the narrow scope of electronic Floquet band engineering that rapidly decohere once the light-periodic drive ends.
Instead, long-lived lattice modulations act as an effective, dynamic switch that periodically drives the electronic Hamiltonian where strong electronic correlations amplify lattice nonlinearities into an {\em extreme manifold} via a coherent feedback loop between the electronic and phononic sectors.
This {\em coherence-imprinted hybrid order} can be definitively characterized by distinctive fingerprints of high-order quantum correlations, sustained within the periodically driven phase, which remain fundamentally unreachable in standard lattice systems or equilibrium phases.


The prototypical correlated electron system Ta$_2$NiSe$_5$~\cite{Sunshine1985, Salvo1986, LuNC2017} provides an ideal platform to investigate electronic correlation-enhanced nonlinear phononics. Although its equilibrium phase classification of exciton condensation remains actively debated and is distinct from the non-equilibrium regime studied here, its universally recognized ingredients--including a narrow-gap electronic structure, strong interband Coulomb correlations (excitonic instability), and exceptional exciton-phonon coupling--provide the ideal platform for exciton-amplified nonlinear phononics~\cite{Werdehausen2018,Mazza2020,Subedi2020,Fukutani2021,Baldini2023,Haque2024,YJiang2024,Larkin2018,Windgatter2021}. This highly susceptible electronic bath functions as a non-equilibrium amplifier, imprinting electronic-scale correlations onto driven lattice coordinates to amplify otherwise weak lattice nonlinearities.  
These unique properties establish Ta$_2$NiSe$_5$ as a definitive platform for generating an {\em extreme manifold} of multi-order quantum pathways. This represents a conceptually distinct control mechanism--driven by a highly susceptible non-equilibrium electronic bath--that transcends the limitations of conventional equilibrium, linear, and previously reported nonlinear phononic regimes~\cite{Disa2021, Takamura2024,LuoNP2022}.
The underlying, coherence-imprinted hybrid ``exciton–phonon" effective order is responsible for the correlation-boosted multi-quantum and cross-mode anharmonic phonon-mixing pathways, originating from a {\em periodically driven electronic Hamiltonian by nonlinear phonons}, never previously observed in any standard lattice materials.

Here, we apply THz two-dimensional coherent spectroscopy (THz-2DCS) to resonantly drive infrared (IR)-active phonon modes in Ta$_2$NiSe$_5$ and 
detect correlation-boosted nonlinear phononics in the coherence-imprinted hybrid order. 
THz-2DCS~\cite{HuangNRP2025,ReimannJCP2021,Lu2017,LuoNP2022} serves simultaneously as a selective driver of phonon modes and as a coherence-resolved, multi-correlation probe of the resulting coherence-imprinted, exciton--phonon order response--analogous to how collective excitations reveal superconducting coherence~\cite{LuoNP2022,HuangSA2025,LiuNatPhys2024}. 
Phase-locked, intense THz pulse pairs and 2D mapping of coherent spectra reveal a remarkably diverse landscape of approximately 30 distinct multi-order quantum pathways. This extreme nonlinear phononic manifold features up to three-quantum coherences, fourth-harmonic generation, and six-wave mixing--all emerging from the sophisticated cross-mode anharmonic interactions.
In this sense, THz-2DCS functions as a {\em coherence tomography} tool for the nonlinear phonon-driven quantum state, capable of resolving multi-order correlations in it that are invisible to equilibrium probes ~\cite{HuangNRP2025,ReimannJCP2021,XLiPRL2006,AGSalvadorPRB2024}. 
These nonlinear phononic signals collapse with increasing temperature near $\sim$100~K, tracking the electronic correlation/coherence scale of the hybrid exciton--phonon order that sustains these high-order quantum pathways. 
Our quantitative simulations demonstrate that purely lattice-mechanical models underpredict the observed nonlinear signal sizes by orders of magnitude and fail to capture the complex multi-correlation pathways resolved, proving that the extreme degree of nonlinear phononics requires Coulomb-mediated amplification from the material’s highly susceptible electronic bath.

\subsubsection*{Correlation-boosted nonlinear phononics driven by phase-locked THz pulses}\label{sec2}

\begin{figure} 
		\centering
		\includegraphics[width=0.8\textwidth]{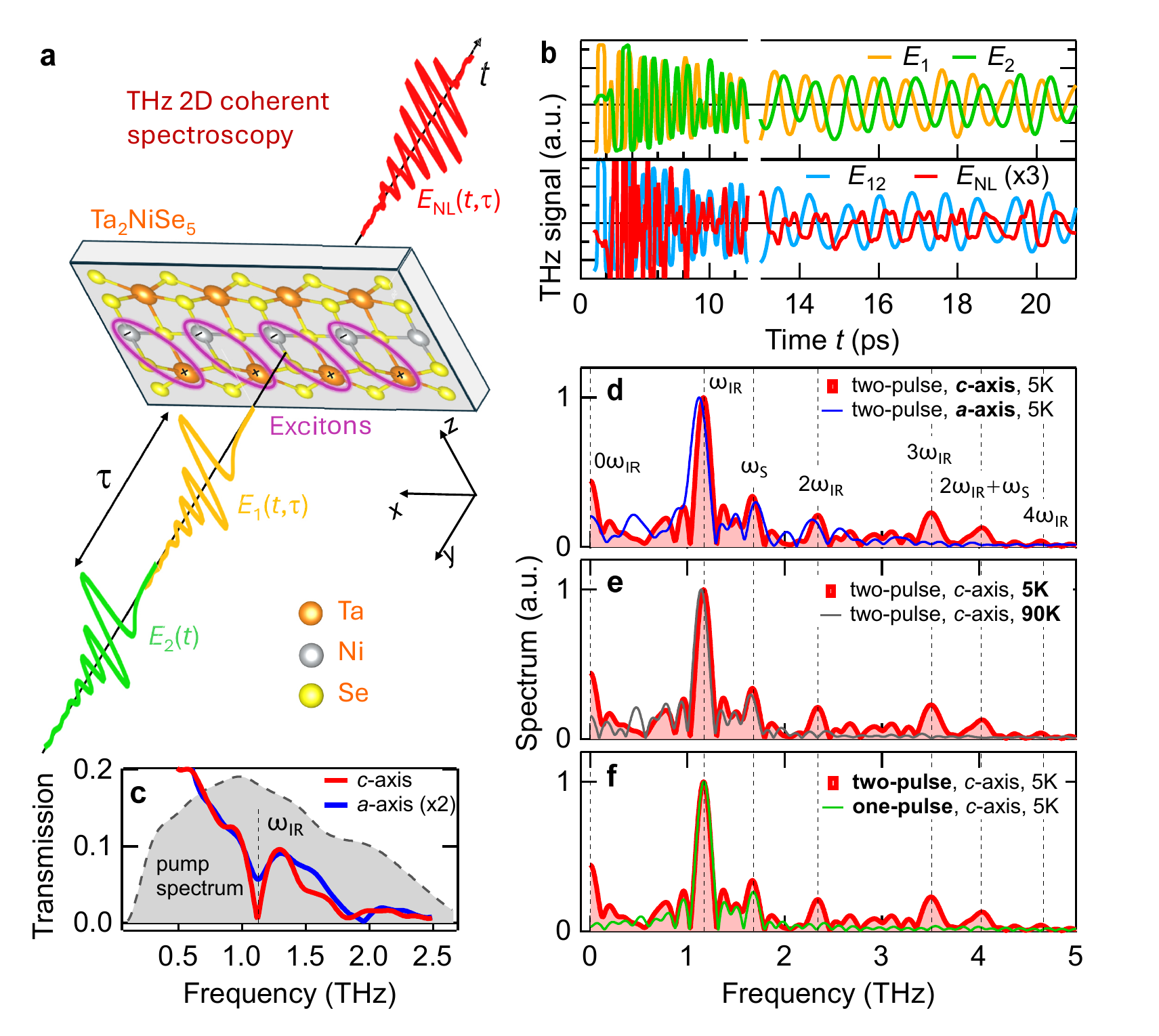} 
		
		\caption{\textbf{Nonlinear phonon generation and emission driven by intense THz pulse pairs.}
			\textbf{a}, Schematic of the THz-2DCS experiment. Two phase-locked  THz pulses $E_1(t, \tau)$ (yellow) and $E_2(t)$ (green) with an interpulse delay $\tau$ are focused onto the sample. Nonlinear THz emission from the sample $E_\text{NL}(t, \tau)$ (red) is measured. \textbf{b}, Representative four channels of THz signal ($E_{1}$, $E_{2}$, $E_{12}$, and $E_\text{NL}$) measured at inter-pulse delay $\tau=2$~ps and $T=5$~K. 
            \textbf{c}, Spectrum of the intense THz pulse centered near $\sim 1$~THz (gray) and static THz transmission of the sample measured at $T=5$~K with THz probe along $a$- and $c$-axes (blue and red). 
            $\textbf{d}$--$\textbf{f}$, The spectrum $E_\text{NL}(\omega_t,\tau)$ under two-pulse excitation at $\tau=2$~ps, $T=5$~K, and along $c$-axis (all red curves), obtained via Fourier transformation of $E_\text{NL}$ from (b) over the time window 14--21~ps. To highlight the effects of varying excitation conditions, additional normalized spectra are plotted for comparison: (d) excitation along $a$-axis (blue), (e) at $T=90$~K (gray), and (f) under one-pulse excitation by $E_2$ only (green), while other conditions remain the same for each panel. The vertical dashed lines mark the position of 0$\omega_\text{IR}$, $\omega_\text{IR}$, $\omega_\text{S}$, 2$\omega_\text{IR}$, 3$\omega_\text{IR}$, 2$\omega_\text{IR}$+$\omega_\text{S}$, and 4$\omega_\text{IR}$, respectively.}
		\label{fig:Fig1} 
	\end{figure}

The experimental configuration is shown in Fig.~\ref{fig:Fig1}a (see Methods for details). The nonlinear THz emission, defined as $E_\text{NL}(t, \tau)=E_{12}(t, \tau)-E_1(t, \tau)-E_2(t)$ (red), is extracted from the transmitted THz signals measured with pulse 1 ($E_{1}$, yellow), pulse 2 ($E_{2}$, green), and both pulses ($E_{12}$, blue) present, as shown in Fig.~\ref{fig:Fig1}b. Here, $t$ and $\tau$ denote the real time and the interpulse delay, respectively. 
The THz pulse pair used for excitation is centered at $\sim 1$~THz with a spectral width spanning $\sim 0.3-2.5$~THz (gray, Fig. ~\ref{fig:Fig1}c), which enables resonant excitation of the IR-active B$_\text{1u}$ phonon mode, $Q_\text{IR}$, at $\omega_\text{IR}\sim 1.17$~THz along $c$-axis, as shown by static THz transmission measurement (red, Fig.~\ref{fig:Fig1}c). The same mode appears significantly weaker and broadened, yet still visible, along $a$-axis (blue, Fig.~\ref{fig:Fig1}c). The measured $\omega_\text{IR}$ frequency and transmission anisotropy are consistent with prior reports~\cite{Larkin2018,Takamura2024}.  

Figure~\ref{fig:Fig1}b shows representative time-domain signals of $E_1$, $E_2$, $E_{12}$, and $E_\text{NL}$ up to 21~ps.
In addition to the initial complex signals arising from various electronic and optical nonlinearities, the long-time traces exhibit clear oscillations and beating patterns from phonons. To highlight this nonlinear phononic behavior, the spectrum of $E_\text{NL}$, obtained by Fourier transforming (FT) the time-domain signal $E_\text{NL}(t, \tau)$ from Fig.~\ref{fig:Fig1}b over the 14-21~ps time window, is used to isolate coherent phonon nonlinearities induced by the pulse pairs, shown as red curves in Figs.~\ref{fig:Fig1}d-f. Several well-resolved spectral peaks are observed. To elucidate the salient features of nonlinear phononic modes driven by THz pulse pairs, we perform control measurements under varied conditions--including excitation along $a$-axis (Fig.~\ref{fig:Fig1}d, blue), elevated temperature at $T=90$~K (Fig.~\ref{fig:Fig1}e, gray), and single-pulse excitation (Fig.~\ref{fig:Fig1}f, green)--while keeping all other parameters identical for each panel. All spectra are normalized to their main phonon peak to enable direct comparison across conditions. These comparisons reveal that, aside from the primary $Q_\text{IR}$ phonon mode at $\omega_\text{IR}\sim 1.17$~THz and a secondary IR mode $Q_\text{S}$ near $\omega_\text{S}\sim 1.68$~THz that appear under all excitation conditions, several additional peaks emerge predominantly under two-pulse excitation at $T=5$~K along $c$-axis. 
The spectral peaks at 0, 2.34, 3.51, and 4.68~THz correspond respectively to phononic THz rectification ($0\omega_\text{IR}$) as well as second-, third-, and fourth-harmonic generation ($2\omega_\text{IR}$, $3\omega_\text{IR}$, and $4\omega_\text{IR}$). 
The nonlinear order of these harmonics is confirmed by the THz field-dependent scaling of their amplitudes (Fig.~\ref{fig:Fig2}), showing $E^2$, $E^2$, $E^3$, and $E^4$ scaling for the 0, 2.34, 3.51, and 4.68~THz modes.
Moreover, several weaker side peaks emerge adjacent to the $\omega_\text{IR}$ harmonics, with the most prominent feature appearing at $\sim 4.02$~THz, consistent as a sum-frequency sideband $2\omega_\text{IR}+\omega_\text{S}$.
The nonlinear order of this peak exhibits an $E^3$ scaling behavior  (Fig.~\ref{fig:Fig2}d), corroborating it as a third-order cross-mode anharmonic coupling, $2Q_\text{IR}-1Q_\text{S}$,
rather than a response from a single IR phonon excitation. 


It is also worth emphasizing three key points in Fig.~\ref{fig:Fig1}. First, we resolve strong even-order nonlinear responses (e.g., second- and fourth-harmonic generation, $2\omega_\text{IR}$ and $4\omega_\text{IR}$ in Fig.~\ref{fig:Fig1}d) despite centrosymmetric equilibrium symmetry of conventional lattice. In addition, the spectral bandwidth of the THz driving pulses is limited to below $\sim 2.5$~THz (gray, Fig.~\ref{fig:Fig1}c), mostly resonantly exciting the $Q_\text{IR}$ and $Q_\text{S}$ modes and particularly ruling out direct excitation of any fundamental phonon modes above the range. Consequently, the observed peaks at $3\omega_\text{IR}$, $2\omega_\text{IR}+\omega_\text{S}$, and $4\omega_\text{IR}$ clearly originate from non-resonant, high-order nonlinear processes. 
The symmetry-based discriminator and the strong cross-mode anharmonic mixing suggest the involvement of electronic correlations under THz driving acting as an effective coupling and amplification bath for the nonlinear phononic response.
Second, when the excitation polarization is changed from $c$- to $a$-axis (Fig.~\ref{fig:Fig1}d), the lower-order peaks at $0\omega_\text{IR}$ and $2\omega_\text{IR}$ become noticeably weaker, while the higher-order peaks at $3\omega_\text{IR}$, $2\omega_\text{IR}+\omega_\text{S}$, and $4\omega_\text{IR}$ are almost suppressed. 
This anisotropic behavior reflects the stronger response of the dominant $Q_\text{IR}$ mode along $c$-axis compared to $a$-axis, consistent with THz transmission spectra (Fig.~\ref{fig:Fig1}c) and the IR phonon eigenvector analysis (Fig.~\ref{fig:Fig4}). 
While excitonic instability are delocalized along the a-axis chains, the robust $Q_\text{IR}$ phonon serves as both the driver and reporter of the hybrid phonon-exciton state, reflecting the efficient resonant THz coupling to its large c-axis dipole moment for enable the hybrid response.
Third, these high-frequency nonlinear phononic peaks are absent under single-pulse excitation but emerge prominently under two-pulse coherent excitation (Fig.~\ref{fig:Fig1}f). The comparison 
between single- and two-pulse excitation conditions 
underscores the unique capability of THz-2DCS to drive and detect multi-correlation phenomena that remain inaccessible to conventional single-pulse or one-axis spectroscopic methods.
Consequently, these findings motivate a detailed investigation of temperature tuning, multidimensional coherent control, and multi-correlation pathways in the phonon periodically driven states of Ta$_2$NiSe$_5$, to elucidate the underlying nature of correlation-enhanced nonlinear phononics.



\subsubsection*{Temperature-tuned nonlinear phononics trace the electronic coherence scale of the hybrid order}\label{sec3}
As shown in Fig.~\ref{fig:Fig1}e, raising the temperature from 5~K to 90~K markedly suppresses the nonlinear phononic features, including the high-harmonic generation (HHG) peaks ($2\omega_\text{IR}$, $3\omega_\text{IR}$, and $4\omega_\text{IR}$) and the cross-mode sum-frequency sideband feature ($2\omega_\text{IR}+\omega_\text{S}$). Notably, the linewidths of the resonant $Q_\text{IR}$ and $Q_\text{S}$ phonon peaks exhibit no clear broadening between 5~K and 90~K, consistent with the relatively weak temperature sensitivity of conventional phonon behaviors of these modes in the equilibrium phase. 
By comparison, the dramatic temperature-dependent collapse of the off-resonant nonlinear phononic peaks--contrasting sharply with the persistence of resonant phonon signals--is inconsistent with a purely phononic or structural origin and instead points to a coherence scale associated with the driven dynamical state.
This behavior indicates that the highly susceptible electronic sector plays an essential role in amplifying and sustaining the extreme nonlinear phononic manifold.
If standard lattice mechanics were the driver, these linear modes would evolve strongly. Together, the observation of linear–nonlinear decoupling supports a periodic Hamiltonian engineering framework in Ta$_2$NiSe$_5$: a coherence-imprinted hybrid exciton--phonon order established by utilizing periodic nonlinear lattice modulations as a dynamic switch for the correlated electronic sector. The THz-2DCS multi-correlation signals track the electronic coherence scale rather than equilibrium structural order.

\begin{figure} 
		\centering
		\includegraphics[width=0.7\textwidth]{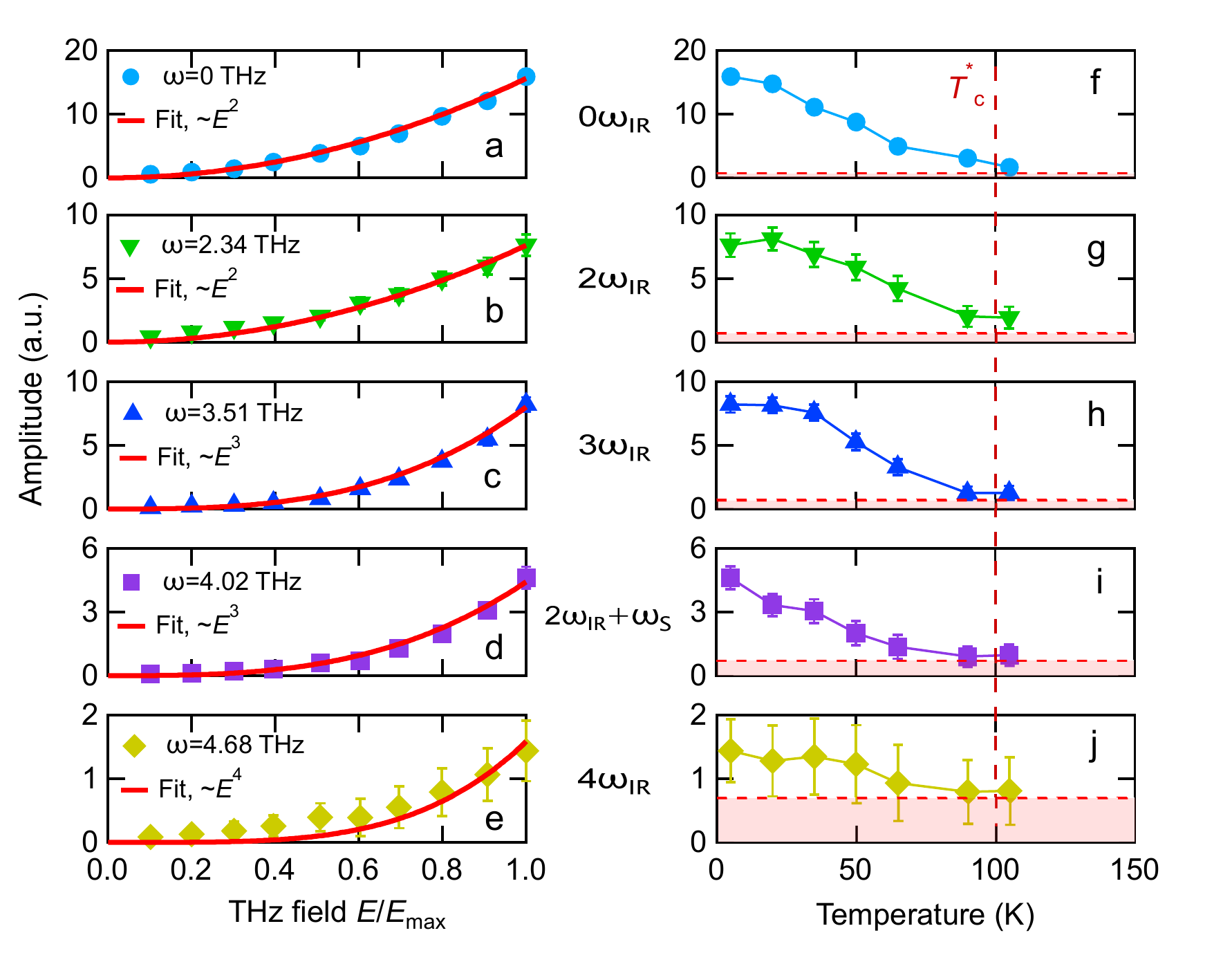} 		
		
	\caption{\textbf{THz field- and temperature-dependent nonlinear responses of exciton-boosted phonons.}
			\textbf{a}--\textbf{e}, Integrated spectral weight (ISW) of off-resonance harmonic and sideband peaks at $0\omega_\text{IR}$, $2\omega_\text{IR}$, $3\omega_\text{IR}$, $2\omega_\text{IR}+\omega_\text{S}$, and $4\omega_\text{IR}$, respectively, plotted as a function of THz field strength. Solid red lines represent theoretical power-law dependencies $\sim E^2$, $E^2$, $E^3$, $E^3$, and $E^4$, respectively, for comparison with the experimental data. Error bars are determined from the ISW of noise floors above 5~THz. \textbf{f}--\textbf{j}, Corresponding ISW plotted as a function of temperature that diminishes at an electronic correlation/coherence scale near $T^{*}_\text{c} \sim 100$~K (vertical red dashed line) of the nonlinear phonon imprinted hybrid exciton--phonon order. The results are obtained from temperature-dependent spectra $E_\text{NL}(\omega)$ recorded at a fixed interpulse delay of $\tau=2$~ps, with the data adjusted for the temperature-dependent transmission of THz pulses through the sample to account for THz transmission changes at different temperatures. Horizontal red dashed lines and shades mark the uncertainty thresholds; spectral peaks below these levels cannot be reliably resolved.}
		\label{fig:Fig2} 
	\end{figure}

To put the physical picture on a firmer footing, 
Figs.~\ref{fig:Fig2}f-j show the integrated spectral weight (ISW) of off-resonant harmonic and sideband peaks at $0\omega_\text{IR}$, $2\omega_\text{IR}$, $3\omega_\text{IR}$, $2\omega_\text{IR}+\omega_\text{S}$, and $4\omega_\text{IR}$, respectively, as a function of temperature. 
All peaks are strongly suppressed with increasing temperature, defining a correlation/coherence scale $T^{*}_{\text{c}}\sim 100$~K. Above $T^{*}_{\text{c}}$ the nonlinear phononic signals fall below the FT uncertainty threshold (red shading) and become unresolvable. We emphasize that $T^{*}_{\text{c}}$ represents a coherence/correlation scale governing the survival of the high-order nonlinear manifold, rather than a thermodynamic transition. Below $T^{*}_{\text{c}}$ electronic correlations can strongly amplify lattice nonlinearities, beyond what is inferred from linear phonon responses and without being restricted to the chain $a-$axis.
Moreover, the strong temperature-dependent collapse of off-resonant nonlinear phonon peaks is absent in optical pumping~\cite{YJiang2024,Werdehausen2018} and resonant phonon response at $\omega_\text{IR}\sim 1.17$~THz (Fig.~\ref{fig:Fig1}e).
These observations support THz-driven, coherence-imprinted hybrid order as the origin of the high-order nonlinear phonon manifold.
As the electronic coherence of the dynamic order weakens with increasing temperature, the correlated background loses its ability to amplify phononic nonlinearities, leading to the observed collapse of high-order correlations and quantum pathways.

\subsubsection*{Multi-correlation tomography and quantum pathways of the coherence-imprinted hybrid order}\label{sec4}

\begin{figure*}[t]
    \centering
		\makebox[\linewidth][c]{\includegraphics[width=.7\linewidth]{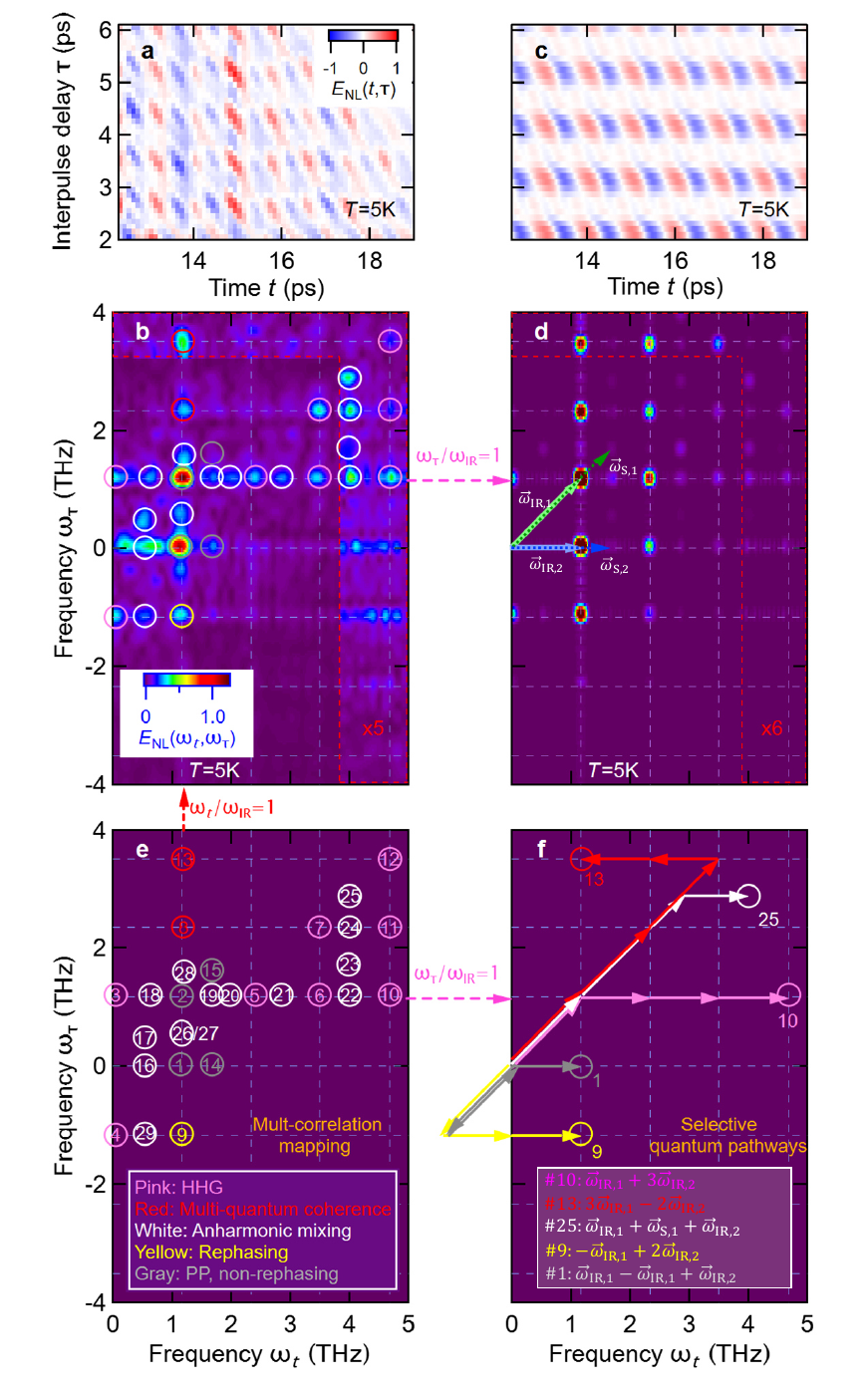}}
    \caption{ \textbf{Multi-order correlation mapping of the hybrid coherence-imprinted exciton--phonon order by THz-2DCS.}
    \textbf{a}, 2D false-color plot of the measured $E_\text{NL}(t,\tau)$ at $T=5$~K along $c$-axis.
\textbf{b}, The corresponding 2D spectrum $E_\text{NL}(\omega_t,\omega_\tau)$ obtained by 2D FT of (a). The horizontal and vertical dashed blue lines mark the frequencies corresponding to multiples of $\omega_\text{IR}$. The region enclosed by the red dashed box is scaled for better visualization of  weaker peaks. Distinct nonlinear peaks are marked by colored circles: pink for high-harmonic generation, red for multi-quantum coherence, white for anharmonic mixing, yellow for rephasing, and gray for pump--probe and non-rephasing signals. 
\textbf{c--d}, Corresponding theoretical calculations of $E_\text{NL}(t,\tau)$ and $E_\text{NL}(\omega_t,\omega_\tau)$, respectively. The solid light green and light blue arrows represent the frequency vectors ${\vec\omega_\text{IR,1}}$ and ${\vec\omega_\text{IR,2}}$, respectively. The dashed dark green and dark blue arrows represent the frequency vectors ${\vec\omega_\text{S,1}}$ and ${\vec\omega_\text{S,2}}$, respectively. \textbf{e}, Schematic of multi-correlation pathways in the coherence-imprinted hybrid order. This dynamic phase sustains exciton-boosted, extreme nonlinear phononics under periodic Hamiltonian driving simulated in (c)-(d).
Each nonlinear peak marked by a colored circle in (b) is labeled with a number for discussion in the text and summarized in Extended Data Table~\ref{tab1}.
\textbf{f}, Representative excitation pathways illustrated for HHG (No.~10), multi-quantum coherence (No.~13), anharmonic mixing (No.~25), rephasing (No.~9) and pump--probe (PP) (No.~1). Inset shows frequency-vector construction for selected modes. 
    }
    \label{fig:Fig3}
\end{figure*}

Figure~\ref{fig:Fig3} reveals the hallmark of the coherence-imprinted hybrid order: an unprecedented density of nonlinear phononic pathways far exceeding the limits of conventional lattice anharmonicity~\cite{LinPRL2022}. Rather than a collection of isolated modes, the observed extreme manifold--comprising approximately 30 distinct peaks--serves as a comprehensive coherence fingerprint of the driven state. This landscape, resolved via THz-2DCS, reveals high-harmonic generation, multi-quantum coherences, and complex cross-mode cascades--ranging from sum-, difference-, and sideband-frequency mixing, to multi-stage up-/down-conversion cascades and single- through quad-quantum manifold couplings. The sheer order and intensity of these pathways provide the central evidence for correlation-amplified nonlinear phononics: while the centrosymmetric lattice possesses intrinsically weak, odd-only nonlinearities, the highly susceptible excitonic instability acts as a non-equilibrium amplifier, imprinting electronic-scale correlations onto the nonlinear phonon responses.

We employ full THz-2DCS by scanning both the emission time $t$ and interpulse delay $\tau$. The 2D false-color map of $E_\text{NL}(t,\tau)$ at $T = 5$~K along $c$-axis (Fig.~\ref{fig:Fig3}a) displays strong oscillations along both excitation and emission time axes. 
These long-lived oscillations trace the temporal trajectory of the coherence-imprinted hybrid order, demonstrating that the extreme nonlinear phononic manifold and its multicomponent correlations persist after the light is turned off, well beyond the duration of the initial THz excitation.
Oscillatory features along the vertical axis $\tau$ further reveal coherent control of these nonlinear responses, resulting from constructive and destructive interference between phase coherences induced by the two THz excitations. 

The corresponding 2D spectrum $E_\text{NL}(\omega_t,\omega_\tau)$, obtained from 2D FT of $E_\text{NL}(t,\tau)$, is shown in Fig.~\ref{fig:Fig3}b. It reveals an exceptionally rich array of well-resolved spectral peaks--circled and labeled Nos.~1-29 in Fig.~\ref{fig:Fig3}e--far surpassing nonlinear features previously reported in nonlinear phononics, where observation and analysis were constrained to 1D spectra.
To illustrate their identifications, we first define two frequency vectors, ${\vec\omega}_\text{IR,1}=(\omega_\text{IR}, \omega_\text{IR})$ and ${\vec\omega_\text{IR,2}}=(\omega_\text{IR}, 0)$, as marked by solid light green and light blue arrows in Fig.~\ref{fig:Fig3}d, corresponding to the $Q_\text{IR}$ phonon driven by THz pulses 1 and 2, respectively~\cite{HuangNC2024,LuoNP2022,HuangSA2025}. 
With these notations, peaks Nos.~1–2 correspond to the fundamental pump--probe and nonrephasing pathways at $(\omega_\text{IR},0)$ and $(\omega_\text{IR},\omega_\text{IR})$, respectively, while peak No.~9 represents the rephasing (R) process at $(\omega_\text{IR},-\omega_\text{IR})$, characteristic of coherent echo-type responses of the $Q_\text{IR}$ mode. 
Each peak in Fig.~\ref{fig:Fig3}b thus maps a distinct multi-order quantum pathway within the coherence-imprinted hybrid exciton--phonon order, resolving an exceptionally rich landscape of approximately 30 distinct processes (Extended Data Table~1), as elaborated below. 

First, peaks at the intersections of the horizontal and vertical dashed lines correspond to anharmonic mixing among $\omega_\text{IR}$ manifolds arising solely from the $Q_\text{IR}$ mode. These include THz phonon rectification (Nos.~3 and~4) at $(0, \pm\omega_\text{IR})$, second-harmonic generation 
(No.~5) at $(2\omega_\text{IR}, \omega_\text{IR})$, third-harmonic generation 
(Nos.~6 and~7) at $(3\omega_\text{IR}, \omega_\text{IR})$ and $(3\omega_\text{IR}$, 2$\omega_\text{IR})$, and fourth-harmonic generation via high-order coherence mixing of $Q_\text{IR}$ manifolds 
(Nos.~10-12) at $(4\omega_\text{IR}, \omega_\text{IR})$, $(4\omega_\text{IR}, 2\omega_\text{IR})$, and $(4\omega_\text{IR}, 3\omega_\text{IR})$.
Moreover, each anharmonic mixing peak above can be schematically built in the 2D plane, such as peak No.~10 (pink, Fig.~\ref{fig:Fig3}f) with an excitation pathway $1{\vec\omega}_\text{IR,1} + 3{\vec\omega}_\text{IR,2}$.
Some of these peaks, shown as merged features in the 1D spectra presented in Fig.~\ref{fig:Fig1}, are now disentangled through 2D mapping of quantum pathways, where their multi-photon excitation pathways are resolved. 
For example, the peak No.~12 at $(4\omega_\text{IR}, 3\omega_\text{IR})$ arises from absorption of three photons from pulse 1 and one photon from pulse 2, i.e., $(4\omega_\text{IR}, 3\omega_\text{IR})=3{\vec\omega}_\text{IR,1}+{\vec\omega}_\text{IR,2}$. 

Second, more intriguingly, the 2D spectrum unveils additional multi-quantum nonlinear phononic excitation processes that are hidden in 1D emission spectra, such as the two-quantum coherence (2QC, No.~8) at $(\omega_\text{IR}, 2\omega_\text{IR})=2{\vec\omega}_\text{IR,1}-{\vec\omega}_\text{IR,2}$ and the three-quantum coherence (3QC, No.~13) at $(\omega_\text{IR}, 3\omega_\text{IR})=3{\vec\omega}_\text{IR,1}-2{\vec\omega}_\text{IR,2}$. These features appear along the excitation frequency axis~$\omega_{\tau}$ \cite{Rohrbach2025}, different from the high harmonic generation peaks that emerge along emission frequency axis~$\omega_{t}$~\cite{TurnrNature2010,SYuOL2019,vonHoegen2018}. They manifest as 2QC–1Q and 3QC–1Q down-conversion cascade cross-peaks in the 2D plane. For example, the 3QC-1Q peak (No.~13) results from THz pulse 1 interacting three times with the $Q_\text{IR}$ mode to generate a three-quantum phonon coherence, which subsequently interacts twice with THz pulse 2 in opposite phase to yield a single-quantum phonon coherence at the observed position illustrated in Fig.~\ref{fig:Fig3}f. A detailed assignment of nonlinear processes involving only  the $Q_\text{IR}$ mode is summarized in Extended Data Table~\ref{tab1} (Nos.~1--13).

\clearpage  

Third, remarkably, beyond the peaks arising solely from $Q_\text{IR}$ manifolds, a rich set of peaks alongside the $\omega_\text{IR}$ harmonics, associated with the $Q_\text{S}$ mode (Nos.~14 and 15) and sideband-modulated cross-mode peaks (Nos.~16–29), is uniquely resolved in the coherence-imprinted hybrid order regime. 
A closer inspection reveals that most of the numbered features can be represented by $n\omega_\text{IR}\pm\omega_\text{S}$ (where $n$ is an integer), such as the four peaks at $\omega_t\sim$4.02 THz (Nos.~22-25), pointing to anharmonic coupling between $Q_\text{IR}$ and $Q_\text{S}$ modes. To identify these processes, we introduce another pair of frequency vectors for the $Q_\text{S}$ mode, i.~e., $\vec{\omega}_\text{S,1}=(\omega_\text{S}$, $\omega_\text{S}$) and $\vec{\omega}_\text{S,2}=(\omega_\text{S}$, 0), driven by THz pulse 1 and 2, respectively, as marked by dashed dark green
and dark blue arrows in Fig.~\ref{fig:Fig3}d. Consequently, the four peaks above can be obtained by $(2\omega_\text{IR}+\omega_\text{S}, \omega_\text{IR})=\vec{\omega}_\text{IR,1}+\vec{\omega}_\text{IR,2}+\vec{\omega}_\text{S,2}$, $(2\omega_\text{IR}+\omega_\text{S}, \omega_\text{S})=\vec{\omega}_\text{S,1}+2\vec{\omega}_\text{IR,2}$, $(2\omega_\text{IR}+\omega_\text{S}, 2\omega_\text{IR})=2\vec{\omega}_\text{IR,1}+\vec{\omega}_\text{S,2}$, and $(2\omega_\text{IR}+\omega_\text{S}, \omega_\text{IR}+\omega_\text{S})=\vec{\omega}_\text{IR,1}+\vec{\omega}_\text{S,1}+\vec{\omega}_\text{IR,2}$. 
All of these sum-frequency sideband cross-peaks exhibit cubic anharmonic phonon mixing, consistent with THz field dependent measurement shown in Fig.~\ref{fig:Fig2}d. 
A detailed assignment of nonlinear processes involving $Q_\text{S}$ mode is provided in Extended Data Table~\ref{tab1} and also elaborated in Supplementary Note~3. 
The exceptional cross-mode anharmonic mixing exemplifies the extreme multi-correlations and nonlinear interference that underlie the coherence-imprinted, hybrid exciton–phonon order, phenomena largely absent in other lattice systems~\cite{Ghalgaoui2025,Blank2023,LinPRL2022} and even surpassing THz magnonic platforms~\cite{HuangNC2024,ZhangNatPhys2024}.

\subsubsection*{Discussion}\label{sec5}


We perform a model calculation of the underlying dynamics based on the coherent coupling between phonon modes and the excitonic instability, utilizing a microscopic framework adapted from Refs.~\cite{MurakamiPRL2017,Millis2020} (see Supplementary Note~1).
This model serves as a theoretical validation of the essential mechanism where the highly susceptible electronic sector acts as a non-equilibrium amplifier to sustain the extreme nonlinear phononic response that pure lattice modes alone fail to capture (see Supplementary Note~4).
The simulation results are presented in Figs.~\ref{fig:Fig3}c and \ref{fig:Fig3}d, which show the calculated nonlinear signal $E_\text{NL}(t,\tau)$ and $E_\text{NL}(\omega_t,\omega_\tau)$, respectively. The simulations qualitatively capture the key experimental features (Figs.~\ref{fig:Fig3}a and \ref{fig:Fig3}b), including high harmonic generation and multi-quantum coherence peaks associated with the dominant $Q_\text{IR}$ mode, as well as its anharmonic coupling with the weaker $Q_\text{S}$ mode that mainly appears as sidebands. 
This agreement confirms that the fundamental origin of the extreme nonlinear manifold is the coherent coupling between phonons and excitonic instabilities, where the large-amplitude $Q(t)$ lattice coordinate acts as an effective, phononic ``Floquet" switch for the periodically temporal modulation of the electronic Hamiltonian. This periodic Hamiltonian modulation imprints lattice coherence onto the electronic sector, sustaining an extreme manifold of multi-quantum correlations that purely mechanical lattice models fail to reproduce (see Supplementary Note~4). 
Minor quantitative differences between experiment and theory are expected given the simplified nature of the model, yet they do not affect the central picture of exciton-boosted nonlinear phononics and instead motivate more advanced theoretical simulations and refined parameter estimates.

\begin{figure}[] 
	\centering
	\includegraphics[width=.9\textwidth]{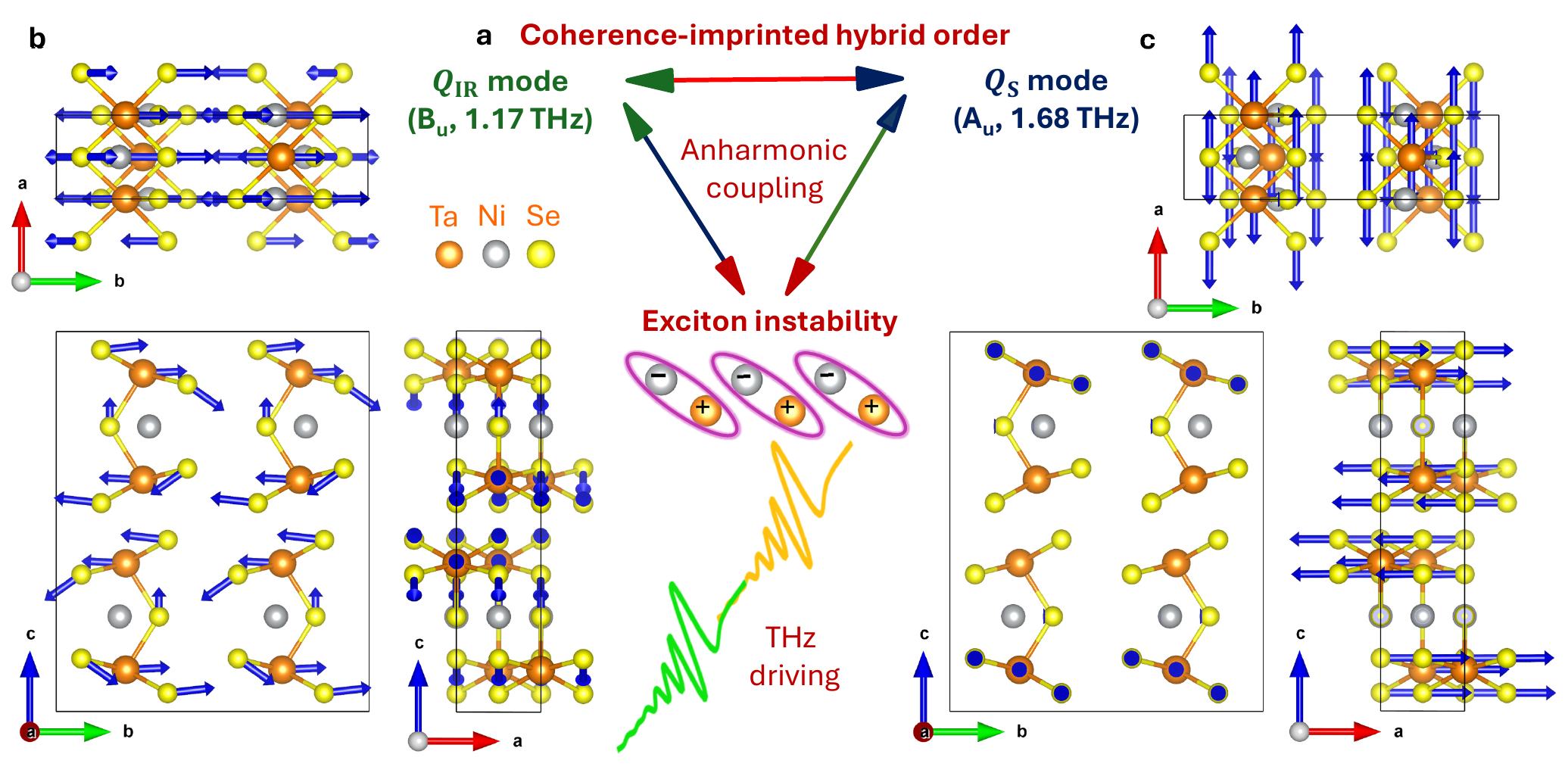} 
    \caption{
    \textbf{Phonon eigenvectors of two low-frequency IR modes in Ta$_2$NiSe$_5$ and their anharmonic coupling.} 
    \textbf{a}, A schematic illustrating the anharmonic coupling between $Q_\mathrm{IR}$ and $Q_\mathrm{S}$ phonon modes and excitons. \textbf{b}, Side and top views of the $Q_\mathrm{IR}$ mode. The displacement of Ta and Se atoms induces a dipole moment along both $a$- and $c$-axes. Ta, Ni, and Se atoms are shown in orange, gray, and yellow, respectively. The conventional monoclinic unit cell is outlined, and crystallographic axes are indicated for reference.
    \textbf{c}, Side and top views of the $Q_\mathrm{S}$ mode. The dominant shear motion of the Ta-Se layers along $a$-axis induces a dipole moment predominantly along $b$-axis. Given the C$_\mathrm{2h}$ point-group symmetry of the equilibrium structure, anharmonic cross-mode coupling terms such as $Q_\mathrm{IR}^3Q_\mathrm{S}$ are symmetry-allowed and contribute to the nonlinear mode mixing.
    With the dynamical breaking of inversion symmetry, as suggested by the experimentally observed second-harmonic generation, cubic anharmonic coupling terms like $Q_\mathrm{IR}^2 Q_\mathrm{S}$ become symmetry-allowed. Together with exciton--phonon coupling, they contribute to the observed exciton-boosted multi-order phonon correlations.
    }
	\label{fig:Fig4}
\end{figure}

To gain further material-specific insights into the relevant phonon modes and their anharmonic interactions, we performed first-principles calculations based on density functional theory (DFT). The calculated bulk phonon spectrum reveals two IR–active optical modes within the experimentally relevant frequency range of 1–2 THz, consistent with prior study~\cite{Werdehausen2018} (see Supplementary Note~2), as shown in Fig.~\ref{fig:Fig4}.

The first mode, with B$_\text{u}$ symmetry, appears at a calculated frequency of 1.07~THz. By symmetry, it possesses a dipole moment in the $ac$ plane, with a larger component along $c$-axis as indicated by the eigen-mode shown in Fig.~\ref{fig:Fig4}b,
consistent with experimental observations (Fig.~\ref{fig:Fig1}c). We thus assign it to the experimentally observed $Q_\text{IR}$ mode at 1.17~THz.
The second mode, with A$_\text{u}$ symmetry, appears at a calculated frequency of 1.35 THz. 
Note that although this mode is dominated by the shear motion of the Ta-Se layers along $a$-axis as shown in Fig.~\ref{fig:Fig4}c, the displacement-induced dipole moment lies predominantly along $b$-axis due to A$_\text{u}$ symmetry. This explains why static THz transmission spectra show no detectable signal from $Q_\text{S}$ mode along either $a$- or $c$-axis (Fig.~\ref{fig:Fig1}c), consistent with previous static THz absorption measurements of Ta$_2$NiSe$_5$ crystals in $ac$ plane~\cite{Larkin2018,Takamura2024}. Numerically, we confirm that the mode effective charge for the $Q_\text{S}$ mode has a nonzero $b$-component, while that of $Q_\text{IR}$ has nonzero $a$- and $c$-components, in full agreement with the experimental results and symmetry analysis. We therefore attribute it to the experimentally observed $Q_\text{S}$ mode. Although the $Q_\text{S}$ mode exhibits weak IR activity in the $ac$ plane, it emerges under intense single-pulse THz excitation, and its amplitude is further enhanced under two-pulse coherent excitation (Fig.~\ref{fig:Fig1}f). Owing to its relatively weak IR activity in the $ac$ plane, the $Q_\text{S}$ mode primarily manifests as sideband features, distinct from the strong response of the $Q_\text{IR}$ mode.




In the centrosymmetric equilibrium structure, the lowest-order anharmonic coupling arises from a quartic interaction term, $Q_\text{IR}^3 Q_\text{S}$. Under strong THz excitation and exciton-boosted nonlinearities, the inversion symmetry can be dynamically broken, as evidenced by the emergence of second-harmonic generation, thereby allowing additional cubic coupling terms of the form $Q_\text{IR}^2 Q_\text{S}$. To test whether such lattice anharmonic interactions alone can account for the observed nonlinear response, we performed numerical simulations of the coupled $Q_\text{IR}$ and $Q_\text{S}$ phonon modes with the potential energy surface expanded up to the fourth order (see Supplementary Note~4). Under inversion-symmetric conditions, the allowed quartic phonon--phonon couplings produce nonlinear responses that are two to five orders of magnitude too weak compared to the fundamental mode. Even when strong cubic anharmonicities are introduced to emulate complete dynamical symmetry breaking, the resulting high-harmonic intensities remain far below the experimental observations, and the key multi-quantum pathways remain absent from the calculated 2D spectra. There is no strong even-order response or extreme phonon quantum manifolds (Figure~3). Furthermore, purely lattice anharmonic models underpredict the nonlinear signals by orders of magnitude (see Supplementary Note~4), which is insufficient to explain the experimentally observed nonlinear 2D coherent spectra of Ta$_2$NiSe$_5$. Consequently, electronic correlation amplification under THz driving plays a key role in accounting for the observed giant multi-order nonlinear response. 
In addition, the symmetry mismatch and linear-nonlinear coupling between the equilibrium and dynamic states  (see Supplementary Note~5) is exceptional, providing another defining feature of the coherence-imprinted hybrid order. These observations force the exciton--phonon coupling into higher-order nonlinear channels, directly aligning with our observation of strong even-odd order nonlinearities and an extreme manifold of multi-quantum pathways. Such a mechanism effectively relaxes equilibrium selection rules, allowing the electronic amplifier to boost otherwise forbidden lattice transitions into the resolved $\sim$30 quantum pathways (Fig.~3).

\subsubsection*{Conclusion}\label{sec6}


We establish electronic correlation-amplified nonlinear phononics as a robust principle for multi-order quantum control. By leveraging electronic instabilities as a non-equilibrium amplifier, we resolve an extreme manifold of $\sim$30 quantum pathways, which establishes a new benchmark for THz nonlinear phononics, as the extreme degree of quantum pathways resolved exceeds conventional lattice responses and even state-of-the-art THz magnonics. This coherence-imprinted hybrid order is sustained by phonon-anchored, periodic Hamiltonian engineering, overcoming the rapid decoherence upon light-off for conventional electronic Floquet engineering with periodic drive. By introducing THz-2DCS as a multi-correlation tomography tool, we provide a general framework for certifying periodically driven quantum matter, bridging the fundamental gap between nonlinear phononic control and light-induced phase transitions.



\setcounter{table}{1}
\renewcommand{\thetable}{S\arabic{table}}

\vspace{15pt}

\renewcommand{\tablename}{Extended Data Table}
\renewcommand{\thetable}{\arabic{table}}
\setcounter{table}{0}

\begin{table} 
		\centering
		\caption{\textbf{Nonlinear processes contributing to the THz 2D spectra.} Spectral peaks are summarized from Fig.~\ref{fig:Fig3}b. Peaks Nos.~1-13 (Nos.~14, 15) arise exclusively from $Q_\text{IR}$ ($Q_\text{S}$) mode manifolds; Nos.~16-29 from anharmonic mixing between $Q_\text{IR}$ and $Q_\text{S}$;  Abbreviations: TR (THz rectification), $n$QC ($n$-quantum coherence), R (rephasing), NR (non-rephasing), PP (pump--probe), SHG/THG/4HG (second-/third-/fourth-harmonic generation), $n$WM ($n$th-order wave mixing). 
		}
		\label{tab1} 
		
		\begin{small}	
			\renewcommand{\arraystretch}{1.2}
			\begin{tabular}{|c|c|c|c|} 
					\hline
					No. & 2D Label & Nonlinear Pathways & Coordinates (Emission, Excitation) [THz] \\ 
					\hline
					1 & PP & ${\vec\omega_\text{IR,1}} - {\vec\omega_\text{IR,1}} + {\vec\omega_\text{IR,2}}$ & $(\omega_\text{IR}, 0)$, (1.17, 0) \\
					\hline
					
					2& NR & ${\vec\omega_\text{IR,1}} + {\vec\omega_\text{IR,2}} - {\vec\omega_\text{IR,2}}$ & $(\omega_\text{IR}, \omega_\text{IR})$, (1.17, 1.17)\\
					\hline
					3& TR & ${\vec\omega_\text{IR,1}} - {\vec\omega_\text{IR,2}}$ & $(0, \omega_\text{IR})$, (0, 1.17) \\
					\hline
					
					4& TR & $-{\vec\omega_\text{IR,1}} + {\vec\omega_\text{IR,2}}$ & $(0, -\omega_\text{IR})$, (0, $-$1.17) \\
					\hline
					
					5&	SHG & ${\vec\omega_\text{IR,1}} + {\vec\omega_\text{IR,2}}$ & $(2\omega_\text{IR}, \omega_\text{IR})$, (2.34, 1.17) \\
					\hline
					6&THG & ${\vec\omega_\text{IR,1}} + 2{\vec\omega_\text{IR,2}}$ & $(3\omega_\text{IR}, \omega_\text{IR})$, (3.51, 1.17) \\
					\hline    
					7&	THG & $2{\vec\omega_\text{IR,1}} + {\vec\omega_\text{IR,2}}$ & $(3\omega_\text{IR}, 2\omega_\text{IR})$, (3.51, 2.34) \\
					\hline
					8&	2QC & $2{\vec\omega_\text{IR,1}} - {\vec\omega_\text{IR,2}}$ & $(\omega_\text{IR}, 2\omega_\text{IR})$, (1.17, 2.34) \\
					\hline     
					9&	R & $-{\vec\omega_\text{IR,1}} + 2{\vec\omega_\text{IR,2}} $ & $(\omega_\text{IR}, -\omega_\text{IR})$, (1.17, $-$1.17) \\
					\hline                            
					10&	4HG & ${\vec\omega_\text{IR,1}} + 3{\vec\omega_\text{IR,2}}$ & $(4\omega_\text{IR}, \omega_\text{IR})$, (4.68, 1.17) \\
					\hline   
					
					11&	4HG & $2{\vec\omega_\text{IR,1}} + 2{\vec\omega_\text{IR,2}}$ & $(4\omega_\text{IR}, 2\omega_\text{IR})$, (4.68, 2.34) \\
					\hline
					
					12&  4HG & $3{\vec\omega_\text{IR,1}} + {\vec\omega_\text{IR,2}}$ & $(4\omega_\text{IR}, 3\omega_\text{IR})$, (4.68, 3.51) \\
					\hline            
					13&	3QC & $3{\vec\omega_\text{IR,1}} - 2{\vec\omega_\text{IR,2}}$ & $(\omega_\text{IR}, 3\omega_\text{IR})$, (1.17, 3.51) \\
					\hline \hline
					
					14&	PP & ${\vec\omega_\text{S,1}} - {\vec\omega_\text{S,1}}+{\vec\omega_\text{S,2}}$ & $(\omega_\text{S}, 0)$, (1.68, 0) \\
					\hline        
					15&	NR & ${\vec\omega_\text{S,1}} + {\vec\omega_\text{S,2}}-{\vec\omega_\text{S,2}}$ & $(\omega_\text{S}, \omega_\text{S})$, (1.68, 1.68) \\
					\hline \hline

					16&	5WM & ${\vec\omega_\text{IR,1}} - {\vec\omega_\text{IR,1}}-{\vec\omega_\text{IR,2}}+{\vec\omega_\text{S,2}}$ & $(\omega_\text{S}-\omega_\text{IR}, 0)$, (0.51, 0) \\
					\hline  
					
					17&	5WM & $-{\vec\omega_\text{IR,1}} + {\vec\omega_\text{S,1}} + {\vec\omega_\text{IR,2}}-{\vec\omega_\text{IR,2}}$ & $(\omega_\text{S}-\omega_\text{IR}, \omega_\text{S}-\omega_\text{IR})$, (0.51, 0.51) \\
					\hline 
					
					18&	4WM & ${\vec\omega_\text{IR,1}} + {\vec\omega_\text{IR,2}}-{\vec\omega_\text{S,2}}$ & $(2\omega_\text{IR}-\omega_\text{S}, \omega_\text{IR})$, (0.66, 1.17) \\
					\hline

					19&	4WM & ${\vec\omega_\text{IR,1}} - {\vec\omega_\text{IR,2}}+{\vec\omega_\text{S,2}}$ & $(\omega_\text{S}, \omega_\text{IR})$, (1.68, 1.17) \\
					\hline     
					
					20&	5WM & ${\vec\omega_\text{IR,1}} + 2{\vec\omega_\text{IR,2}}-{\vec\omega_\text{S,2}}$ & $(3\omega_\text{IR}-\omega_\text{S}, \omega_\text{IR})$, (1.83, 1.17) \\
					\hline 
					21&	5WM & ${\vec\omega_\text{IR,1}} + {\vec\omega_\text{IR,2}}-{\vec\omega_\text{IR,2}}+{\vec\omega_\text{S,2}}$ & $(\omega_\text{IR}+\omega_\text{S}, \omega_\text{IR})$, (2.85, 1.17) \\
					\hline 
					
					22&	4WM & ${\vec\omega_\text{IR,1}} + {\vec\omega_\text{IR,2}}+{\vec\omega_\text{S,2}}$ & $(2\omega_\text{IR}+\omega_\text{S}, \omega_\text{IR})$, (4.02, 1.17) \\
					\hline 
					
					23&	4WM & $ {\vec\omega_\text{S,1}}+2{\vec\omega_\text{IR,2}}$ & $(2\omega_\text{IR}+\omega_\text{S}, \omega_\text{S})$, (4.02, 1.68) \\
					\hline 
					
					24&	4WM & ${2\vec\omega_\text{IR,1}} + {\vec\omega_\text{S,2}}$ & $(2\omega_\text{IR}+\omega_\text{S}, 2\omega_\text{IR})$, (4.02, 2.34) \\
					\hline 
					
					25&	4WM & ${\vec\omega_\text{IR,1}} + {\vec\omega_\text{S,1} + {\vec\omega_\text{IR,2}}}$ & $(2\omega_\text{IR}+\omega_\text{S}, \omega_\text{IR}+\omega_\text{S})$, (4.02, 2.85) \\
					\hline 
					
					26&	6WM & ${2\vec\omega_\text{IR,1}} -{\vec\omega_\text{S,1}} - {\vec\omega_\text{IR,2}} + {\vec\omega_\text{S,2}}$ & $(\omega_\text{IR}, 2\omega_\text{IR}-\omega_\text{S})$, (1.17, 0.66) \\
					\hline 
					
					27&	6WM & $-{\vec\omega_\text{IR,1}} + {\vec\omega_\text{S,1}} + 2{\vec\omega_\text{IR,2}} - {\vec\omega_\text{S,2}}$ & $(\omega_\text{IR}, \omega_\text{S}-\omega_\text{IR})$, (1.17, 0.51) \\
					\hline 
					
					28&	4WM & ${\vec\omega_\text{S,1}} - {\vec\omega_\text{S,2}} + {\vec\omega_\text{IR,2}}$ & $(\omega_\text{IR}, \omega_\text{S})$, (1.17, 1.68) \\
					\hline

					29&	3WM & $-{\vec\omega_\text{IR,1}} +{\vec\omega_\text{S,2}}$ & $(\omega_\text{S}-\omega_\text{IR}, -\omega_\text{IR})$, (0.51, -1.17) \\
					\hline

				\end{tabular}
				\renewcommand{\arraystretch}{1.0}
			\end{small}
		\end{table}

\newpage

\bibliography{ref}

\backmatter

\bmhead{Data availability} 
The data needed to evaluate the conclusions of the paper are provided in the main text and the supplementary information. Data for all figures presented in this study are available on DataShare. 

\bmhead{Acknowledgements}
The THz spectroscopy study was supported by the U.S. Department of Energy, Office of Basic Energy Science, Division of Materials Sciences and Engineering (Ames National Laboratory is operated for the U.S. Department of Energy by Iowa State University under Contract No. DE-AC02-07CH11358).
The research by M.M., T.J. and Y.-X.Y. used resources of the National Energy Research Scientific Computing Center (NERSC), a Department of Energy User Facility. The work at BIT was supported by the National Key Research and Development Program of China (Grant No. 2022YFA1403400), the Beijing Natural Science Foundation (Grant No. Z210006), and the Beijing National Laboratory for Condensed Matter Physics (Grant No. 2023BNLCMPKF007). Z.W. thanks the Analysis and Testing Center at BIT for assistance in facility support. 

\bmhead{Author contributions}
L.L. and J.W. conceived the project. L.L. and A.K. performed THz spectroscopic experiments. L.Y., Z.W., and Y.Y. synthesized and characterized the sample. M.M. and Y.-X.Y. performed quantum kinetic simulations. T.J. and Y.-X.Y. performed DFT calculations. L.L., M.M., Y.-X.Y., and J.W. analyzed data with the help of A.K., T.J., Z.C., C.H., Z.X.C., J.P., I.E.P., and D.X. The manuscript was written by J.W., L.L., M.M., T.J., and Y.-X.Y. with discussions from all authors. J.W. supervised the project

\bmhead{Competing interests} 
There are no competing interests to declare.


\end{document}